\documentclass{article}
\usepackage[utf8]{inputenc}
\usepackage[a4paper,margin=1.3in]{geometry}
\usepackage{amsmath,dsfont}
\usepackage{amsthm,subcaption,graphicx}
\usepackage{amsfonts,amssymb}
\usepackage[font=small,labelfont=bf]{caption}
\usepackage{authblk}

\usepackage[authoryear,round]{natbib}
\usepackage{dsfont}
\usepackage{subcaption}
\usepackage[font={footnotesize,it}]{caption}
\usepackage{xcolor}
\usepackage{enumitem}
\usepackage{blindtext}
\usepackage{url}
\usepackage{hyperref}
\newtheorem{theorem}{Theorem}

\definecolor{oranje}{cmyk}{0,0.50,0.84,0}

\newcommand{\Var}{\mathrm{Var}}

\newcommand{\argmax}[1]{\underset{#1}{\mathrm{argmax}}\ }
\newcommand{\bs}[1]{\boldsymbol{#1}}

\newcommand{\dd}{\mathrm{d}}
\newcommand{\mc}[1]{\mathcal{#1}}

\usepackage{algorithmic,algorithm}

\title{Fast marginal likelihood estimation of penalties for group-adaptive elastic net}
\author[1]{Mirrelijn M. van Nee\thanks{ 
The first author is supported by ZonMw TOP grant COMPUTE CANCER (40-
00812-98-16012).}}
\author[1]{Tim van de Brug}
\author[1,2]{Mark A. van de Wiel}
\affil[1]{{\small{Epidemiology and Data Science, Amsterdam University Medical Centers, The Netherlands}}}
\affil[2]{{\small{MRC Biostatistics Unit, Cambridge University, UK}}}
\date{}

\begin{document}

\maketitle

\begin{abstract}
Nowadays, clinical research routinely uses omics data, such as gene expression, for predicting clinical outcomes or selecting markers. 
Additionally, so-called co-data are often available, providing complementary information on the covariates, like p-values from previously published studies or groups of genes corresponding to pathways.
Elastic net penalisation is widely used for prediction and covariate selection.
Group-adaptive elastic net penalisation learns from co-data to improve the prediction and covariate selection, by penalising important groups of covariates less than other groups.
Existing methods are, however, computationally expensive.
Here we present a fast method for marginal likelihood estimation of group-adaptive elastic net penalties for generalised linear models. 
We first derive a low-dimensional representation of the Taylor approximation of the marginal likelihood and its first derivative for group-adaptive ridge penalties, to efficiently estimate these penalties.
Then we show by using asymptotic normality of the linear predictors that the marginal likelihood for elastic net models may be approximated well by the marginal likelihood for ridge models. 
The ridge group penalties are then transformed to elastic net group penalties by using the variance function.
The method allows for overlapping groups and unpenalised variables.
We demonstrate the method in a model-based simulation study and an application to cancer genomics. The method substantially decreases computation time and outperforms or matches other methods by learning from co-data.

\end{abstract}

% \begin{keyword}
% \kwd{clinical prediction}
% \kwd{omics}
% \kwd{prior information}
% \kwd{penalised generalised linear models}
% \kwd{empirical Bayes.}
% \end{keyword}

\section{Introduction}
%clinical studies omics for prediction
%use of co-data or features of features to improve prediction and covariate selection
Clinical studies increasingly involve the analysis of omics data, like gene expression, metabolomics and radiomics. The research goal may be to predict some outcome, like therapy response or diagnosis, or selecting markers for follow-up study.  
In addition to the main data with information on the samples, so-called co-data may be available, providing complementary information on the covariates. In cancer genomics, for example, genes are grouped according to pathways or gene ontology, and p-values may be derived from previous studies available in public repositories like The Cancer Genome Atlas \citep{tomczak2015cancer}. 
The predictions and covariate selection may improve by learning from co-data.

%elastic net popular approach as it estimates and selects covariates simultaneously
%existing group-adaptive methods ipflasso proposes to search over a grid of penalties, but computationally intractable for large grids or more than 5 groups. 
%gren proposed variational bayes approach but only for binomial and logistic
A popular approach to prediction and covariate selection is elastic net penalisation \citep{zou2005regularization}, as it simultaneously estimates and selects covariates.
Recent work includes co-data by allowing for differential group penalties, penalising informative groups of covariates less than non-informative ones. 
The method \texttt{fwelnet} \citep{tay2020feature} extends use of grouped co-data to continuous co-data (termed features of features). The method may be used for grouped co-data as well, for which it is shown to correspond to an elastic net penalty on the group level, with the amount of penalisation governed by one global penalty parameter. Though fast, this method may not be flexible enough when groups differ largely in prediction strength.
The method \texttt{ipflasso} \citep{boulesteix2017ipf} selects the best group penalties from a proposed grid of possible values and is therefore able to flexibly adapt group penalties. In general, however, it is unclear what values should be proposed and only a finite set of values may be proposed. Moreover, the method is not scalable in the number of groups as the number of possible grid configurations increases exponentially.
The method \texttt{gren} \citep{munch2018adaptive}, for group-adaptive elastic net, uses a variational Bayes algorithm to compute empirical Bayes estimates for the group penalties in logistic regression. The method is group-adaptive and scalable in the number of groups, but not in the number of covariates. Moreover, it is only implemented for binary response.
Note that group-adaptive methods are very different in nature than (variations of) the group lasso \citep{Meier2008}. The latter selects entire groups. While this may be useful when there are many small groups, its limited flexibility in terms of penalisation may render inferior predictive performance for settings with a limited number of groups \citep{munch2018adaptive}. 

%this paper
Here we propose a fast and efficient method for marginal likelihood estimation of group elastic net penalties. The method is group-adaptive and may be used for elastic net penalised generalised linear models in high-dimensional data. We include details for linear and logistic regression. Groups may be overlapping and the method allows for unpenalised covariates, such as an intercept or clinical covariates like age and sex. First, we derive a low-dimensional representation of the marginal likelihood for ridge models, a special case of elastic net. Then we show by using the asymptotic multivariate normality of the linear predictor that this marginal likelihood also approximates the marginal likelihood of elastic net models well. Lastly, we show how the ridge group penalties are transformed to find optimal marginal likelihood elastic net group penalties. 

%outline
The outline of the paper is as follows.
Section \ref{par:method} includes details of the method. 
Section \ref{par:results} first compares performance and computation times of several methods in a linear regression model-based simulation study. 
After, the method is illustrated in the logistic regression setting on a cancer genomics data example.
Section \ref{par:discussion} then concludes and discusses the method and results.

% Clinical studies 

% Main contributions:
% \begin{itemize}
%     \item Provide fast marginal likelihood optimisation for obtaining multi-group ridge penalty estimates for ridge penalised generalised linear models. Combine results from wood paper with ideas from multiridge paper to obtain Laplace appropximation of the log marginal likelihood and partial derivatives for ridge penalised generalised linear models in high-dimensional data. Approximations are fast as computations are in terms of number of groups instead of number of covariates, plus unpenalised covariates (like an intercept) may be included.
%     \item We use a multivariate central limit theorem to show that the same marginal likelihood approximation may be used for a larger class of priors, including the sparser elastic net prior.
%     \item We provide an R-package for obtaining fast approximate maximum marginal likelihood estimates for group penalties of a multi-group elastic net.
% \end{itemize}

\section{Method}\label{par:method}
Let the response be given by $\bs{Y}\in\mathbb{R}^n$ and the observed high-dimensional data, $p>n$, by $X\in\mathbb{R}^{n\times p}$. 
Let the co-data matrix $Z\in\mathbb{R}^{p\times G}$ contain group membership information for $G$ groups as defined in \citep{vanNee2020flexible}. 
First assume that groups are not overlapping. Below, we show how the method may be used for partly overlapping groups too.
Let the observed data for group $g$ be given by $X_g$.
We model the response with a generalised linear model, and impose a group-regularised elastic net prior on the regression coefficients $\bs{\beta}\in\mathbb{R}^p$, with group penalties $\bs{\lambda}\in\mathbb{R}^G$:
\begin{align}\label{eq:glm}
\begin{split}
    Y_i | \bs{X}_i,\beta &\overset{ind.}{\sim} \pi\left(Y_i | \bs{X}_i,\beta\right),\ \mu_i:=E_{Y_i|\bs{X}_i,\bs{\beta}}(Y_i)=g^{-1}(\bs{X}_i\bs{\beta}),\ i=1,..,n,\\
    \pi(\beta_k|\alpha,\lambda_{g_k})&\propto\exp\left(-\lambda_{g_k}(\alpha|\beta_k|+(1-\alpha)/2\beta_k^2)\right),\ k=1,..,p,
\end{split}
\end{align}
with $g(\cdot)$ the link function corresponding to the type of generalised linear model used, $\Var(y_i)=\phi V(\mu_i)$ for some known variance function $V(\cdot)$ and scale parameter $\phi$, and $\lambda_{g_k}$ the group-specific penalty of group $g$ to which covariate $k$ belongs.

We use an empirical Bayes approach and estimate the group penalties for a given $\alpha$ to arrive at the group-adaptive regularised elastic net estimate for the regression coefficients:
\begin{align}
    \hat{\bs{\lambda}} &= \argmax{\bs{\lambda}} \pi(\bs{Y}|\bs{\lambda},\alpha) = \argmax{\bs{\lambda}} \int_{\bs{\beta}}\pi(\bs{Y}|\bs{\beta})\pi(\bs{\beta}|\bs{\lambda},\alpha)\dd\bs{\beta}, \label{eq:MML1}\\
    \hat{\bs{\beta}} &= \argmax{\bs{\beta}}\left\{ \log(\pi(\bs{Y}|\bs{\beta})) + \log(\pi(\bs{\beta}|\hat{\bs{\lambda}},\alpha))\right\}.
\end{align}
The data $X$ possibly contain some unpenalised variables next to the penalised variables, e.g. for an intercept, which we will refer to by $X_{unpen}\in\mathbb{R}^{n\times p_1}$ and $X_{pen}\in\mathbb{R}^{n\times p_2}$ respectively, $p=p_1+p_2$. In that case, only the penalised variables are integrated out in Equation \eqref{eq:MML1}. We refer to the unpenalised and penalised regression coefficients by $\bs{\beta}_{unpen}$ and $\bs{\beta}_{pen}$ respectively.

\subsection{Fast marginal likelihood estimation for group-adaptive ridge models}
First, consider group-regularised ridge models ($\alpha=0$).
\cite{wood2011fast} derives a Laplace approximation for the marginal likelihood for semiparametric generalised linear models when the prior distribution is of the form:
\begin{align}\label{eq:priorwood}
    \bs{\beta}\sim N(\bs{0},\phi S^{-}),\ S=\sum_g\lambda_gS_{g},
\end{align}
with $S^{-}$ a generalised inverse and $S$ the ridge penalty matrix equal to the weighted sum over $G$ different known penalty matrices $S_g$.
The \texttt{R}-package \texttt{mgcv} implementing this approximation is developed for low-dimensional data and does not allow for high-dimensional data.
In theory, the results include the case of ridge models in high-dimensional data as well, i.e. define $S_g=I_g\in\mathbb{R}^{p\times p}$, with $I_g$ the diagonal matrix with the $k^{th}$ diagonal element equal to $1$ if covariate $k$ belongs to group $g$ and $0$ otherwise. 
In practice, however, the Laplace approximation may be inaccurate for high-dimensional integrals \citep{van2019learning} and is computationally expensive due to the large dimension of $p$.

For ridge models, we may rewrite the high, $p$-dimensional integral as a lower, $n$-dimensional integral by observing that the likelihood only depends on $\bs{\beta}$ via the linear predictors $\bs{\eta}=X\bs{\beta}$ \citep{veerman2019estimation}.
The resulting prior distribution for $\bs{\eta}_{pen}=X_{pen}\bs{\beta}_{pen}\in\mathbb{R}^n$ is again a multivariate normal distribution:
\begin{align}
    \bs{\eta}_{pen}|\bs{\lambda},\phi\sim N\left(0,X_{pen}\phi\Lambda_{pen}^{-1}X_{pen}^T\right)=N\left(0,\phi\sum_g \lambda_{g}^{-1}X_gX_g^T\right),
\end{align}
with $\Lambda_{pen}\in\mathbb{R}^{p_2\times p_2}$ denoting the diagonal ridge penalty matrix for the penalised variables. Note that in general, the penalty matrix $(\sum_g \lambda_{g}^{-1}X_gX_g^T)^{-1}$ cannot be written in the weighted combination of $G$ penalty matrices as in Equation \eqref{eq:priorwood} when $G>1$. 
Moreover, when unpenalised variables are included, the dimension of $(\bs{\beta}_{unpen}^T,\bs{\eta}_{pen}^T)^T$ is still larger than the number of samples, preventing straightforwardly using the existing software for including only one group of high-dimensional data too.

Here, we derive a low-dimensional representation of the high-dimensional Laplace approximated marginal likelihood and its first derivative as derived by \citep{wood2011fast} that may be computed efficiently for multiple groups and allows for inclusion of unpenalised variables.

Recently, it was shown that the maximum likelihood estimate for the linear predictors, $\hat{\bs{\eta}}$, may be obtained efficiently by rewriting the steps of the iterative weighted least squares algorithm (IWLS) in $n$-dimensional terms \citep{van2020fast}. 
We use similar results to rewrite the Laplace approximation and its first derivative in low-dimensional terms. We then use a general purpose optimiser to optimise the marginal likelihood for group-regularised ridge models.
Below we state the results, details are given in Appendix \ref{ap:method}.

The Laplace approximation of the minus log marginal likelihood for group-regularised ridge models is:
\begin{align}\label{eq:MML}
    -\ell(\bs{\lambda},\phi)&\approx -\ell(\hat{\bs{\eta}},\phi) + \frac{1}{2\phi}(\bs{y}-\bs{\mu})^T\hat{\bs{\eta}} -\frac{1}{2}\log(|I_{n\times n} - WH_{pen}|),
\end{align}
with $\ell(\bs{c})$ denoting the log likelihood given parameters $\bs{c}$, $I_{n\times n}\in\mathbb{R}^{n\times n}$ the identity matrix, $W\in\mathbb{R}^{n\times n}$ the weight matrix in IWLS, and $H_{pen}$ the hat matrix for the penalised variables only:
\begin{align}\label{eq:hatPen}
    H_{pen} &= X_{pen}(X_{pen}^TWX_{pen}+\Lambda_{pen})^{-1}X_{pen}^T,
\end{align}
with $\Lambda_{pen}$ the diagonal penalty matrix containing the group penalties.
We use the efficient expression of the $n\times n$-dimensional hat matrix as derived in \citep{van2020fast}.

We state the derivative in terms of $\rho_g=\log(\lambda_g)$.
First we define some useful expressions. 
Let $I_g\in\mathbb{R}^{p\times p}$ denote the diagonal matrix with diagonal element $k$ equal to $1$ if covariate $k$ belongs to group $g$ and $0$ otherwise. 
Denote the full hat matrix by $H$, i.e. as in Equation \eqref{eq:hatPen} but with $X_{pen}$ substituted by $X$, and denote the contribution of the $g^{th}$ group to the hat matrix by $H_g:=X(X^TWX+\Lambda)^{-1}I_gX^T$, with the efficient lower-dimensional representation given in Appendix \ref{ap:method}.
The partial derivatives of the Laplace approximation of the log marginal likelihood to the group parameters are given by, $g=1,..,G$:
\begin{align}\label{eq:partiallam}
\begin{split}
    \frac{\partial -\ell(\bs{\rho},\phi)}{\partial \rho_g} &= \frac{1}{\phi} \left(\bs{y}-\bs{\mu}\right)^TV^{-1}G'^{-1} \left[I_{n\times n} - HW\right] H_g^T\left(\bs{y}-\bs{\mu}+W\hat{\bs{\eta}}\right)\\
   &\qquad -\frac{1}{2\phi}\left(-G'^{-1}\hat{\bs{\eta}} + \bs{y}-\bs{\mu} \right)^T\left[I_{n\times n} - HW\right] H_g^T\left(\bs{y}-\bs{\mu}+W\hat{\bs{\eta}}\right)\\
   &\qquad +\frac{1}{2}tr\left( H_{pen}\frac{\partial W}{\partial\rho_g}\right)  - \frac{1}{2}tr\left(\exp(-\rho_g)X_gX_g^T\left(W^{-1}+\sum_{g=1}^G \lambda_g^{-1}X_gX_g^T\right)^{-1}\right).
\end{split}
\end{align}
where $V\in\mathbb{R}^{n\times n}$ is the diagonal matrix with diagonal elements $V(\mu_i)$, $G'$ is the diagonal matrix with diagonal elements $g'(\mu_i)$, and the partial derivative $\frac{\partial W}{\partial\rho_g}$ readily obtained from \citep{wood2011fast}. The partial derivatives for $\lambda_g$ are obtained by multiplying the derivative to $\rho_g$ by $\frac{1}{\lambda_g}$.

We optimise $\phi$ jointly with the group parameters. As $\hat{\bs{\eta}}$ does not depend on $\phi$, the partial derivative is given by:
\begin{align}\label{eq:partialphi}
    \frac{\partial -\ell(\bs{\rho},\phi)}{\partial \phi} &= \frac{\partial -\ell(\hat{\bs{\eta}},\phi)}{\partial \phi} - \frac{1}{2\phi^2}(\bs{y}-\bs{\mu})^T\hat{\bs{\eta}}.
\end{align}
The partial derivative of the log likelihood depends on the type of glm considered and is known for canonical models. We include details for linear regression in Appendix \ref{ap:method}. For logistic regression, $\phi=1$ is constant and the partial derivative is $0$.

\subsection{Fast marginal likelihood estimation for group-adaptive elastic net models}
Next, we describe how we use the ridge penalty estimates to obtain elastic net penalties with $\alpha\in[0,1]$. Without loss of generality, consider the penalised variables only and leave out subscripts ${\cdot}_{pen}$ for notational convenience. Denote the ridge penalty parameters and penalty matrix by $\bs{\lambda}_R$ and $\Lambda_R$ respectively. 
Define the variance function of the elastic net prior distribution for a given $\alpha\in[0,1]$ as $h(\cdot)$:
%, and define the covariance matrix of $\bs{\beta}$ under the prior as function of the group penalties, $C(\bs{\lambda})$:
\begin{align}
     h(\lambda_{g_k}) &:= \Var_{\bs{\beta}|\alpha,\bs{\lambda}}(\beta_k). %= [\Cov_{\bs{\beta}|\alpha,\bs{\lambda}}(\bs{\beta})]_{kk},\ C(\bs{\lambda}) :=\Cov_{\bs{\beta}|\alpha,\bs{\lambda}}(\bs{\beta}).
\end{align}
Note that for each $\lambda_{g_k}$, $g_k\in\{1,..,G\}$, there exist $\phi,\bs{\lambda}_R$ such that $h(\lambda_{g_k})=\phi\lambda_{R,g}^{-1}$.
% Note that the covariance matrix $\Cov_{\bs{\beta}|\alpha,\bs{\lambda}}(\bs{\beta})$ under the prior is diagonal and contains maximally $G$ different diagonal elements corresponding to the groups. 
% So, there exist $\phi,\bs{\lambda}_R$ such that $\Cov_{\bs{\beta}|\alpha,\bs{\lambda}}(\bs{\beta})=\phi\Lambda_R^{-1}$ %and an inverse $h^{-1}(\cdot)$ such that $h^{-1}(\phi\lambda_{R,g}^{-1})=\lambda_g$.

The prior distribution for $\bs{\eta}=X\bs{\beta}$ is not analytical for $\alpha>0$. An important observation is that we can exploit the high-dimensionality of the data to approximate the prior for $\bs{\eta}$ with a multivariate normal distribution. The following theorem follows from the multivariate central limit theorem for linear random vector forms in \citep{eicker1966multivariate}:
\begin{theorem}\label{thm:MVNgroups}
Suppose $\beta_j\overset{ind.}{\sim}\pi_{g_j}(\beta_j)$ for $j=1,\ldots,p$ and group-specific prior $\pi_{g_j}(\cdot)$, with $E(\beta_j)=0$ and $\Var(\beta_j)=\tau_{g_j}^2\in(0,\infty)$. For $g=1,\ldots,G$, let $G_g$ be the group size and let $X_g \in \mathbb{R}^{n\times G_g}$ be the weights corresponding to group $g$. Let $\bs{x}_{*j}$ denote the $j^{th}$ column of $X\in \mathbb{R}^{n\times p}$.  Suppose $\bs{x}_{*j} \not= \bs{0} \in \mathbb{R}^n$ for all $j$, $rank(X)=n$ for all $p$, and for $p\to\infty$,
\begin{equation}\label{condition1star}
\max_{j=1,\ldots,p} \bs{x}_{*j}^T \big(X X^T\big)^{-1} \bs{x}_{*j} \to 0.
\end{equation}
Then, for fixed $G$, fixed $n$, and $p\rightarrow\infty$,
\begin{align}
\textstyle \left( \sum_g \tau_g^2 X_g X_g^T \right)^{-1/2} X\bs{\beta} \overset{d}{\rightarrow} N(0, I_{n\times n}),
\end{align}
where $I_{n\times n}$ is the $(n\times n)$-dimensional identity matrix and $\big( \sum_g \tau_g^2 X_g X_g^T \big)^{-1/2}$ is the inverse of the unique positive definite square root of $\sum_g \tau_g^2 X_g X_g^T$.
\end{theorem}
If $n=1$ then condition \eqref{condition1star} is equivalent to 
\[
\max_{j=1,\ldots,p} x_{1j}^2 \Big/ \textstyle \sum_{j=1}^p x_{1j}^2 \to 0,
\]
for $p\to\infty$. Informally, condition \eqref{condition1star} can be interpreted as each variable being asymptotically negligible in size compared to the full data set.
In practice, this condition will be reasonable for most omics, especially as omics data are often standardised, but counter examples may exist such as mutation data with rare mutations.

Hence, the prior on $\bs{\eta}$ under an elastic net prior on $\bs{\beta}$ may be approximated by the following multivariate normal distribution:
\begin{align}
    \pi(\bs{\eta}|\alpha,\bs{\lambda})\approx N\left(0,\sum_g h(\lambda_g)X_gX_g^T\right)=N\left(0,\sum_g \phi \lambda_{R,g}^{-1}X_gX_g^T\right)=\pi(\bs{\eta}|\phi,\bs{\lambda}_R).
\end{align}
Then the marginal likelihood may be approximated as follows:
\begin{align}\label{eq:MMLEN}
    \pi(\bs{Y}|\phi,\alpha,\bs{\lambda}) &=\int_{\bs{\eta}} \pi(\bs{Y}|\bs{\eta},\phi)\pi(\bs{\eta}|\bs{\lambda},\alpha)\dd\eta \approx \int_{\bs{\eta}} \pi(\bs{Y}|\bs{\eta},\phi)\pi(\bs{\eta}|\bs{\lambda}_R,\phi)\dd\eta \nonumber\\
    &= \pi(\bs{Y}|\phi,\bs{\lambda}_R),
\end{align}
where the latter expression is efficiently computed using Equation \eqref{eq:MML}.
Hence, the marginal likelihood of group-regularised elastic net models is approximately the same as that of ridge models, but parametrised differently.
The partial derivatives from the elastic net parametrisation may then be obtained from the partial derivatives for the ridge parametrisation as given in Equations \eqref{eq:partiallam} and \eqref{eq:partialphi} by using the chain rule and a change of variables.

% Furthermore, the partial derivatives may be derived by using the chain rule:
% \begin{align}
%     \frac{\partial \pi(\bs{Y}|\phi,\alpha,\bs{\lambda})}{\partial\lambda_g} &\approx \frac{\partial \pi(\bs{Y}|\phi,\bs{\lambda}_R)}{\partial\lambda_g}= \frac{\partial \pi(\bs{Y}|\phi,\phi^{-1}\bs{\lambda}_R)}{\partial\phi^{-1}\lambda_{R,g}} \cdot \frac{\partial \phi^{-1}\lambda_{R,g}}{\partial \lambda_{g}} \nonumber\\
%     &= \frac{\partial \pi(\bs{Y}|\phi, h(\lambda_g)^{-1})}{\partial h(\lambda_g)^{-1}} \cdot \frac{\partial h(\lambda_g)^{-1}}{\partial \lambda_{g}}.
% \end{align}
% Using a change of variables $\phi=f_1(\phi,\phi^{-1}\bs{\lambda}_R)=\phi$ and $\bs{\lambda}_R=f_2(\phi,\phi^{-1}\bs{\lambda}_R)=\phi\cdot\phi^{-1}\bs{\lambda}_R$ results in:
% \begin{align}
%     \frac{\partial \pi(\bs{Y}|\phi,\alpha,\bs{\lambda})}{\partial\lambda_g} &\approx \left[ \frac{\partial \pi(\bs{Y}|\phi,\bs{\lambda}_R)}{\partial\lambda_{R,g}} \phi \right]\cdot \frac{\partial h(\lambda_g)^{-1}}{\partial \lambda_{g}}\\
%     \frac{\partial \pi(\bs{Y}|\phi,\alpha,\bs{\lambda})}{\partial\phi} &\approx \left[\frac{\partial \pi(\bs{Y}|\phi,\bs{\lambda}_R)}{\partial\phi} + \frac{\partial \pi(\bs{Y}|\phi,\bs{\lambda}_R)}{\partial\bs{\lambda}_R^T}\phi^{-1}\bs{\lambda}_R\right] \cdot \frac{\partial h(\lambda_g)^{-1}}{\partial \lambda_{g}},
% \end{align}
% where the partial derivatives for the ridge parametrisation are given in Equations \eqref{eq:partiallam} and \eqref{eq:partialphi}. 

\begin{figure}
    \centering
    \includegraphics[width=\textwidth]{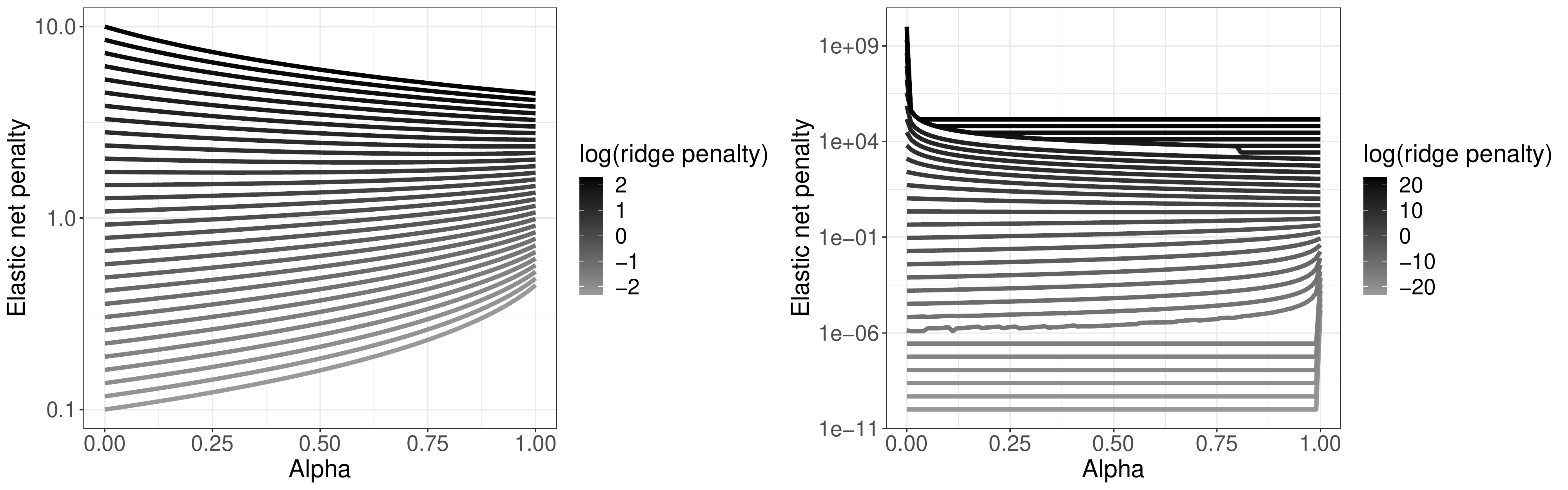}
    \caption{Transformed elastic net penalty for various $\alpha$ and ridge penalty.}
    \label{fig:RtoEN}
\end{figure}

Now, one could again use a general purpose optimiser to maximise the marginal likelihood given in Equation \eqref{eq:MMLEN} for the elastic net parametrisation. 
We use a more direct approach and transform the marginal likelihood estimates for the ridge parametrisation to the elastic net parametrisation using the known variance function.
As the marginal likelihood of the ridge and elastic net model are approximately equal, the maximal value, obtained in the transformed maximiser, is also approximately equal. 
So, the elastic net estimates are given by:
\begin{align}\label{eq:estEN}
    \hat{\bs{\lambda}}&=h^{-1}\left(\hat{\phi}\hat{\bs{\lambda}}_R^{-1}\right),
\end{align}
where $h^{-1}(\cdot)$ is applied element-wise.
% Then these approximately maximise the marginal likelihood, as for all $\bs{\lambda},\phi$ there exist $\bs{\lambda}_R$ such that:
% \begin{align}
%     \pi(\bs{Y}|\hat{\phi},\alpha,\hat{\bs{\lambda}})\approx \pi(\bs{Y}|\hat{\phi},\hat{\bs{\lambda}}_R) \geq \pi(\bs{Y}|\phi,\bs{\lambda}_R) \approx \pi(\bs{Y}|\phi,\alpha,\bs{\lambda}).
% \end{align}
% The difference between the marginal likelihood on the left-hand side and the true maximal value depends on the accuracy of of the Taylor approximation, which increases with $n$, and on the accuracy of the multivariate normal approximation, which increases for increasing $p$ and the number of variables per group. 
% \HL{approximatie geeft mogelijk biased estimates als de approximatie zelf systematisch de ML lager inschat voor sommige waarden dan andere. Hier reCV erin fietsen?}
The proposed approach has the advantage that, once the optimal ridge penalties are obtained, the optimal elastic net penalties are quickly obtained for a whole range of possible $\alpha$ values. Figure \ref{fig:RtoEN} illustrates this for elastic net penalties.
%Below, we describe how to find the solution to Equation \eqref{eq:estEN} for the optimal elastic net penalties.

%\subsubsection{Details elastic net variance function}
The variance function for the elastic net prior for a given $\alpha$ can be shown to be given by:
\begin{align}\label{eq:varEN}
    h(\lambda_g) = \Var_{\bs{\beta}|\alpha,\lambda_g}(\beta_k) 
    = \lambda^{-1}_g(1-\alpha)^{-1} -\frac{\alpha}{\lambda^{1/2}_g(1-\alpha)^{3/2}} \frac{\varphi\left(\frac{\lambda^{\frac{1}{2}}_g\alpha}{(1-\alpha)^{\frac{1}{2}}}\right)}{1-\Phi\left(\frac{\lambda^{\frac{1}{2}}_g\alpha}{(1-\alpha)^{\frac{1}{2}}}\right)} + \frac{\alpha^2}{(1-\alpha)^2},
\end{align}
with $\Phi(\cdot)$ and $\varphi(\cdot)$ the standard normal cumulative density function and probability density function respectively.

The expression simplifies for lasso ($\alpha=1$) to $h(\lambda_g)=\frac{2}{\lambda_g^2}$, and is easily solved analytically.
For $0<\alpha<1$, we use the root-finding algorithm of the \texttt{R}-function \texttt{uniroot} to find the roots of $h(\hat{\phi}\hat{\lambda}_{R,g}^{-1})-\lambda_g=0$.
This suffices for values of $10^{-6}<\lambda_g<10^6$. The evaluation of the variance function is numerically unstable for more extreme values of $\lambda_g$. Therefore we truncate the values to either the ridge or lasso estimate while maintaining the monotonicity for more extreme $\lambda_g$, as illustrated in Figure \ref{fig:RtoEN}.
We expect that this fix will suffice as the precise absolute value has less impact in the extremes.

\subsection{Overlapping groups}
So far, we have assumed the groups to be non-overlapping. A simple way to allow for partly overlapping groups is to make artifical, non-overlapping groups, similarly as proposed as naive implementation of the latent overlapping group lasso \citep{jacob2009group}.

First consider group-regularised ridge models, for which we define $\bs{\tau}=\phi\bs{\lambda}_R^{-1}$. We model the prior variance by the average over multiple groups, given in $Z\bs{\tau}$, as used in \citep{vanNee2020flexible}.
Let $\bar{X}$ denote the observed data matrix where each column $k$ of the original matrix $X$ is duplicated $I_k$ times for each of the $I_k$ groups covariate $k$ belongs to, with group indices given in $\mc{I}_k$. 
Let $\bar{X}'$ denote the matrix where the columns are additionally scaled by $\frac{1}{\sqrt{I_k}}$. 
Define the extended vector of artificial regression coefficients by $\bar{\bs{\beta}}$ and the extended vector with scaled artificial regression coefficients by $\bar{\bs{\beta}}'$.
Each column duplicated from the original column $k$ now corresponds to an artificial, independent effect $\beta_{k_g}=\frac{1}{\sqrt{I_k}}\beta_{k_g}'\overset{ind.}{\sim} N(0,\tau_g^2/I_k)$ for $k=1,..,p,$ and $g\in\mc{I}_k$.
The effect of a covariate $k$ is equal to the sum of the contributions of the groups, $\beta_k=\sum_{g\in\mc{I}_k}\beta_{k_g}=\frac{1}{\sqrt{I_k}}\sum_{g\in\mc{I}_k}\beta_{k_g}'$, such that $X\bs{\beta}=\bar{X}\bar{\bs{\beta}}=\bar{X}'\bar{\bs{\beta}}'$.
The prior distribution of $\beta_{k_g}'$ is given by $N(0,\tau_g^2)=N(0,\phi\lambda_{R,g})$. Hence, we can use our proposed method on the scaled, duplicated $\bar{X}'$ to obtain the estimates for the prior parameters $\phi,\bs{\lambda}_R$.
Finally, the variance estimates are pooled by the co-data matrix $Z$ to compute $\bs{\beta}$ with ridge prior variances $Z\phi\bs{\lambda}_R^{-1}$.

For elastic net models, we first transform the ridge prior variances given in $Z\phi\bs{\lambda}_R^{-1}$ to elastic net penalties as described above.

The high-dimension of $\bar{X}$ increases fast for largely overlapping groups. In practice, however, it is not necessary to store this matrix, nor will it increase the computational time as fast, as the computations for the marginal likelihood only require the high-dimensional computation of $G$ $n\times n$-dimensional matrices $\bar{X}'_g\bar{X}'_g{}^T$ once.
Note that one should take care in including highly overlapping groups as this results in highly correlated groups.

\subsection{Recalibration by cross-validation} 
In particular for non-linear models, like logistic regression, we experienced that the components $\lambda_g$ of $\bs{\lambda}$ were estimated well in a relative sense, but less so in the absolute sense. This may be due to the Laplace approximation of the likelihood, which is
rather coarse for binary outcomes, while being exact for the linear model.
Therefore, the predictive performance may benefit from using recalibrated group-penalties: 
$\bs{\lambda}' = \lambda_0 \bs{\lambda}$, 
where $\bs{\lambda}$ are the estimated penalty factors, and $\lambda_0$ is a global rescaling penalty. As $\lambda_0$ is a scalar, it is efficiently 
and easily estimated by cross-validation, e.g. by using \texttt{glmnet} with penalty factors $\bs{\lambda}$.

\section{Data examples}\label{par:results}
%As gren is not implemented for the linear model, it was not included for this comparison.
The method is termed \texttt{squeezy} as it squeezes out some sparsity from dense group-adaptive ridge models. 
We conduct a model-based linear regression simulation study and illustrate the method on a cancer genomics example in the logistic regression setting.
We include the following elastic net models to compare performance and computation time:
\begin{enumerate}[label=\roman*),noitemsep,topsep=0pt]
\item \texttt{EN} (\texttt{glmnet}, \cite{friedman2010regularization}): a co-data agnostic elastic net penalty, with global penalty parameter obtained by cross-validation; 

\item \texttt{fwEN} and \texttt{fwEN (continuous)} (\texttt{fwelnet}, \cite{tay2020feature}): a globally adaptive elastic net penalty on group level for grouped data (\texttt{fwEN}) or elastic net penalty with weights a function of continuous co-data (\texttt{fwEN (continuous)}). Note that we include \texttt{fwelnet (continuous)} only in the cancer genomics data example for which continuous co-data is available; 

\item \texttt{ipf} and \texttt{ipf2} (\texttt{ipflasso}, \cite{boulesteix2017ipf}): a group-adaptive elastic net penalty, with the group penalty factors selected from a grid of possible values. We take the grid where each penalty factor is in $\{1,2\}$ (\texttt{ipf}) or in $\{1,2,4\}$ (\texttt{ipf2}). Note that we only include the computationally expensive \texttt{ipf2} in the model-based simulated data example for comparison of computing times;

\item \texttt{gren} (\texttt{gren}, \cite{munch2018adaptive}): a group-adaptive elastic net penalty, with the group penalty factors obtained by an approximate empirical-variational Bayes framework. Note that \texttt{gren} is not included in the first linear regression example, as it is not implemented for the linear model;

\item \texttt{ecpcEN squeezy} (\texttt{ecpc}, \cite{vanNee2020flexible} and \texttt{squeezy}): a group-adaptive elastic net penalty. The method \texttt{ecpc} provides empirical Bayes moment estimates for group-adaptive ridge penalties. We use \texttt{squeezy} to transform these to elastic net penalties combined with a recalibrated global rescaling penalty;

\item \texttt{squeezy (single)} and \texttt{squeezy (single+reCV)}: a co-data agnostic elastic net penalty, where the global penalty is obtained by transforming the marginal likelihood estimate for the ridge penalty to an elastic net penalty, without or with recalibration of a global rescaling penalty. While \texttt{glmnet} internally standardises linear response, \texttt{squeezy (single+reCV)} does not;

\item \texttt{squeezy (multi)} and \texttt{squeezy (multi+reCV)}: a group-adaptive elastic net penalty using the proposed method, without or with recalibration of a global rescaling penalty.
\end{enumerate}

\subsection{Model-based simulation study}
We simulate training and test sets of observed data $X$ in $10$ correlated blocks of size $p/10$, regression coefficients from a Laplace distribution ($\alpha=1$) with mean zero and group-specific scale parameter $b_{g_k}$, $g_k\in\{1,..,5\}$, for $5$ equally sized groups, and response from a normal distribution:
\begin{align*}
    \bs{x}_{i,b}&\overset{ind.}{\sim}N\left(\bs{0},I_{(p/10)\times(p/10)}+\frac{\rho^2}{(1-\rho)^2}\bs{1}_{(p/10)\times(p/10)}\right),\ i=1,..,n,\ b=1,..,10,\\ 
    \beta_k&\overset{ind.}{\sim} Laplace(0,b_{k_g}),\ k=1,..,p,\ \bs{Y} \sim N(X\bs{\beta},\sigma^2I_{n\times n}),
\end{align*}
with $\rho=0.2$ and $\sigma^2=2$.
We consider three settings of group parameters: i) no groups: the group scale parameters $\bs{b}$ are the same and chosen such that the variance under each group prior is equal to $(0.1,0.1,0.1,0.1,0.1)\cdot \frac{1200}{p}$; ii) weakly informative groups: the group scale parameters are different, but less so than in the third setting. The variances match $(0.141,0.161,0.186,0.336,0.536)\cdot \frac{1200}{p}$; iii) informative groups: groups are different, with variances matching $(0.01,0.05,0.1,0.4,0.8)\cdot \frac{1200}{p}$.

\begin{figure}
    \centering
    \includegraphics[width=0.5\textwidth]{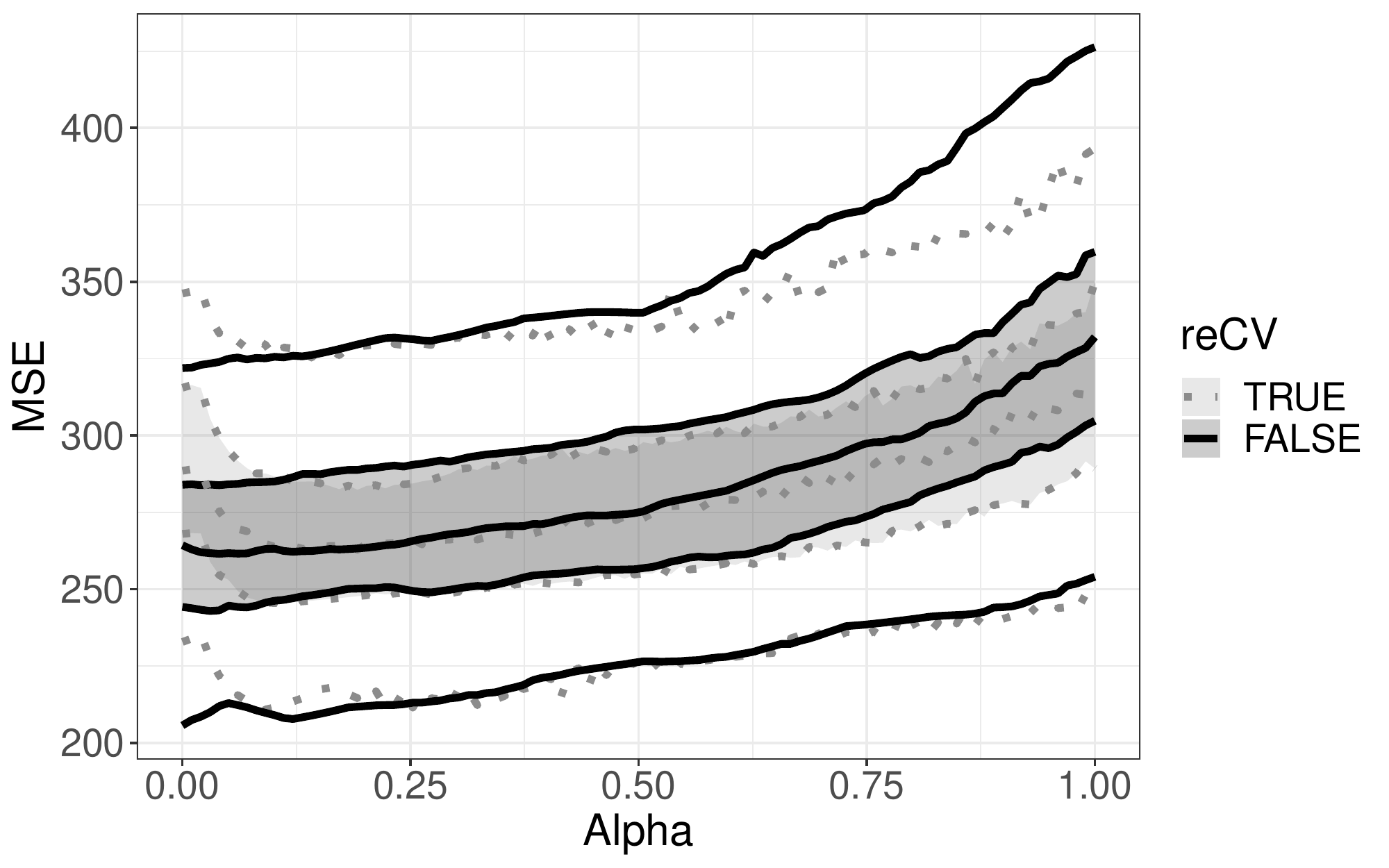}
    \caption{Model-based simulation study, informative groups setting. MSE performance of \texttt{squeezy (multi)} and \texttt{squeezy (multi+reCV)} on the test sets of $100$ pairs of training and test sets for $100$ evenly spaced values of $\alpha\in[0,1]$, shown in pointwise quantiles of $(0.05,0.25,0.5,0.75,0.95)$.}
    \label{fig:SimMSEAlpReCV}
\end{figure}

First, we fit the methods listed above for $\alpha=0.3$ on $100$ independent training and test sets, with $n=150$, $p=600$ and the co-data providing group membership of the $G=5$ groups, to compare performance.
Our method \texttt{squeezy} easily obtains optimal elastic net group penalties for a range of $\alpha$. We observe that there is usually little benefit from recalibration by cross-validation (reCV) in the linear case (Figure \ref{fig:SimMSEAlpReCV}).

\begin{figure}
    \centering
    \includegraphics[width=\textwidth]{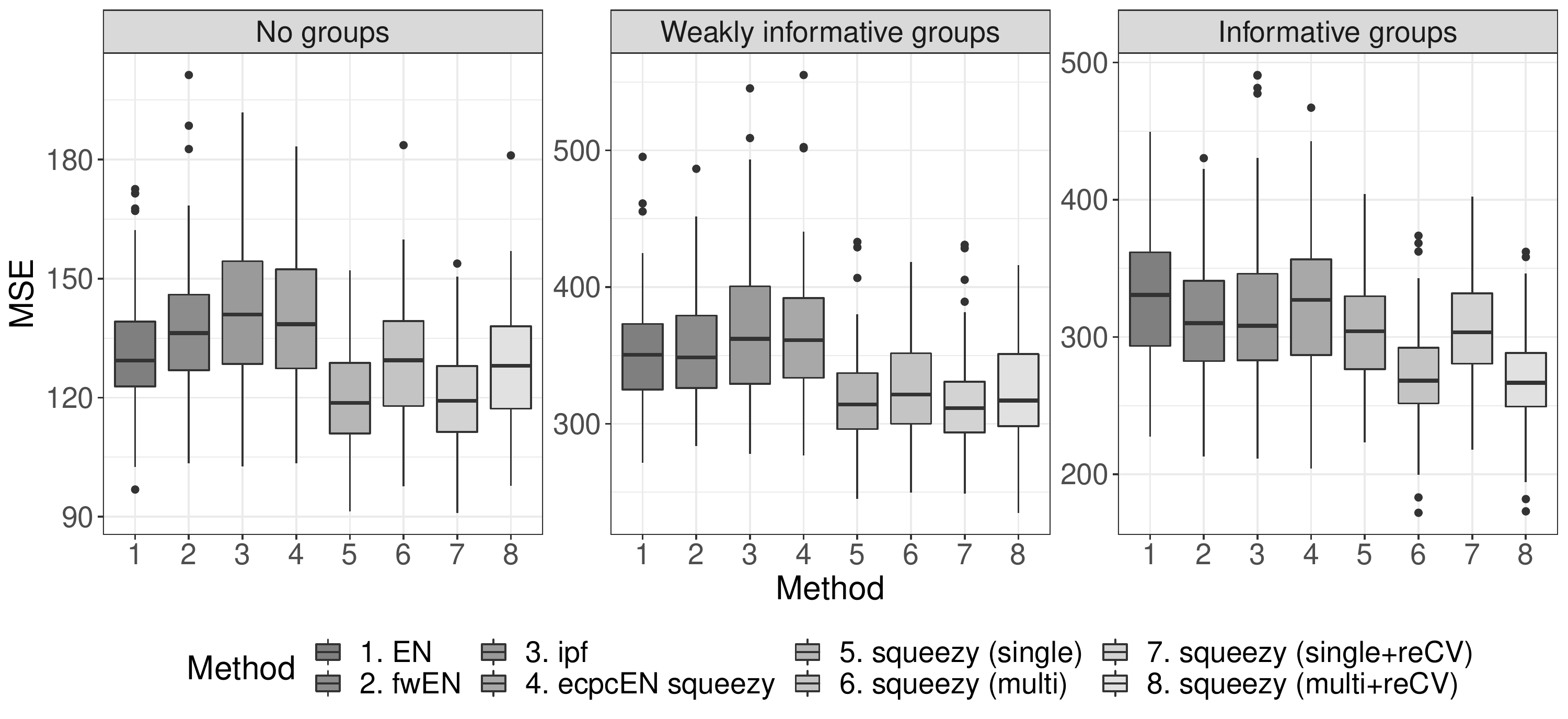}
    \caption{Model-based simulation study. Box plots of the MSE performance on the test sets of $100$ pairs of training and test sets, $\alpha=0.3$.}
    \label{fig:SimMSE}
\end{figure}

\begin{figure}
    \centering
    \includegraphics[width=\textwidth]{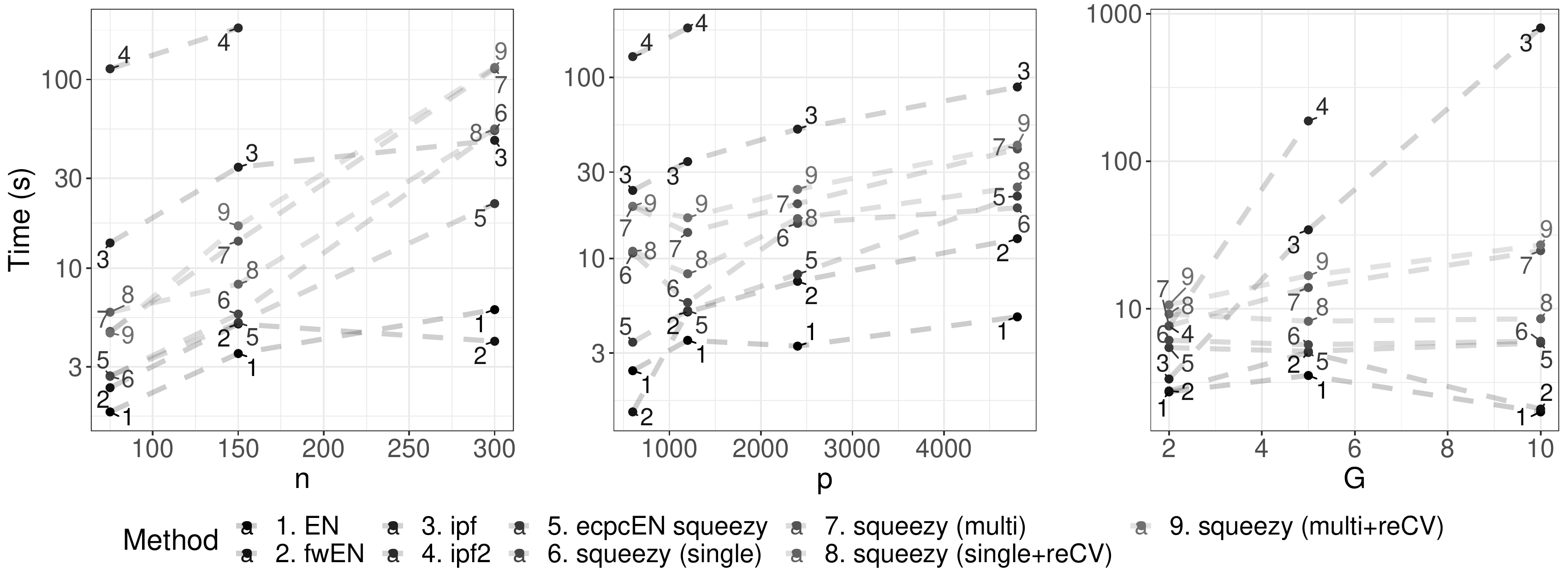}
    \caption{Model-based simulation study. Average computation time in $5$ training sets for varying $n$, $p$ and $G$.}
    \label{fig:SimTime}
\end{figure}

Figure \ref{fig:SimMSE} shows the MSE performance on the test sets for all methods. Our proposed method performs well. The method \texttt{squeezy (single)} (either with or without reCV) outperforms the other methods in the single group setting. The difference in performance between \texttt{squeezy (single+reCV)} and \texttt{EN} may be due to loss of information due to the internally standardisation of the linear response of \texttt{EN}. 
More importantly, \texttt{squeezy (multi)} (either with or without reCV) outperforms the other methods including \texttt{squeezy (single)} in the setting with informative groups, illustrating the benefit of informative co-data. 
The performance of \texttt{squeezy (multi)} and \texttt{squeezy (single)} is more alike in the setting with weakly informative groups, and superior to the other methods.
The method performs better when maximum marginal likelihood estimates for ridge penalties are transformed (\texttt{squeezy}) than when moment estimates are transformed (\texttt{ecpc EN squeezy}). 
Note that the performance of \texttt{ipf} might improve by using a larger grid, but only at a very substantial computational cost as shown in Figure \ref{fig:SimTime}.
While the difference in performance between \texttt{squeezy} and the other methods is smaller for $\alpha=0.8$, \texttt{squeezy (multi)} still outperforms the other methods when the co-data is informative (Figure \ref{fig:SimMSE08} in Appendix \ref{ap:dataexamples}).

Then, we fit the methods on $5$ training sets to compare computation time for varying $n$, $p$ and $G$. The group parameters are set according to the setting with informative groups. We set $(n,p,G)=(150,1200,5)$. Then we vary one of $n,p,G$ while keeping the others fixed.
Figure \ref{fig:SimTime} shows the average computation time for fitting the models.
Unsurprisingly, the group-adaptive methods are slower than the other methods, but \texttt{squeezy} scales substantially better than \texttt{ipf} in the number of groups.

\subsection{Application to predicting therapy response}\label{par:application}
We apply \texttt{squeezy} to predict therapy response (clinical benefit versus progressed disease) using microRNA (miRNA) data from a study on colorectal cancer \citep{neerincx2018combination}, consisting of $p=2114$ miRNA expression levels for $n=88$ independent individuals.
As co-data, we use false discovery rates (FDRs) for differential expression in the primary tumour versus adjacent colorectal tissue. The FDRs were obtained in a previous study \citep{neerincx2015mir} from a different set of non-overlapping samples. 
These co-data were previously shown to be informative for the prediction, miRNAs that are tumor-specific tend to be more important for the response prediction \citep{vanNee2020flexible}.
Here, we discretise the continuous FDRs in $G=8$ equally sized groups, such that we have around $10$ samples per group parameter.

\begin{figure}
    \centering
    \includegraphics[width=\textwidth]{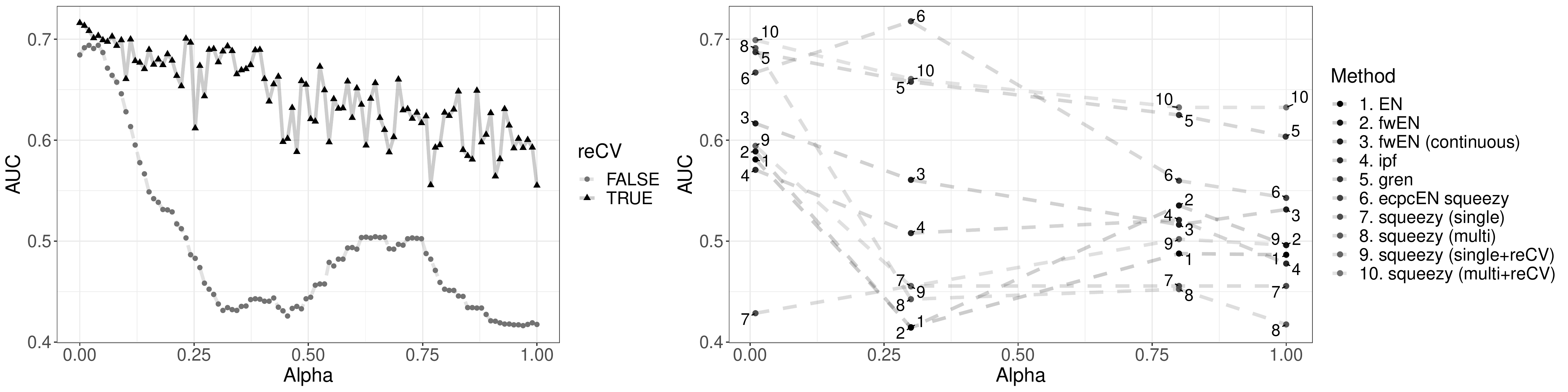}
    \caption{Results of 10-fold cross-validation in miRNA data example. Left: AUC performance for \texttt{squeezy (multi)} and \texttt{squeezy (multi+reCV)} for $100$ evenly spaced values of $\alpha\in[0,1]$. Right: AUC performance for several values of $\alpha$ for various methods.}
    \label{fig:miRNAAUC}
\end{figure}

We fit the methods listed above on $10$ folds to compare cross-validated performance (Figure \ref{fig:miRNAAUC}) and average computation time (Table \ref{tab:miRNATime}).
Figure \ref{fig:miRNAAUC} (left) clearly shows the benefit of recalibration by cross-validation (reCV) in this logistic regression setting. Our method \texttt{squeezy (multi+reCV)} performs as well as \texttt{gren} (Figure \ref{fig:miRNAAUC} (right), cf 10 vs 5) but is around 35 times as fast. The method \texttt{ecpcEN squeezy} is twice as fast as \texttt{squeezy (multi+reCV)} and is competitive to \texttt{squeezy (multi+reCV)} and \texttt{gren}. These three methods all benefit from the informative co-data and outperform the other methods. 

% \begin{figure}
%     \centering
%     \begin{subfigure}[c]{0.45\textwidth}
%     \centering
%     \includegraphics[width=\linewidth]{Figures/squeezy/FigmiRNAAUCalp.pdf}
%     \end{subfigure}
%     \begin{subfigure}[c]{0.54\textwidth}
%     \centering
%     \includegraphics[width=\linewidth]{Figures/squeezy/FigmiRNAAUC.pdf}
%     \end{subfigure}
%     \caption{Results of 10-fold cross-validation in miRNA data example. Left: AUC performance for \texttt{squeezy (multi)} and \texttt{squeezy (multi+reCV)} for $100$ evenly spaced values of $\alpha\in[0,1]$. Right: AUC performance for several values of $\alpha$ for various methods.}
%     \label{fig:miRNAAUC}
% \end{figure}

% \usepackage{graphicx}
\begin{table}[]
\centering
\resizebox{0.9\textwidth}{!}{%
\begin{tabular}{llll}
\hline
\textbf{Method}          & \textbf{Time (s)} & \textbf{Method}          & \textbf{Time (s)} \\ \hline
1. EN                    & 1.17 (1.06)       & 2. fwEN (groups)         & 12.20 (6.84)      \\
7. squeezy (single)      & 5.12 (3.29)       & 10. squeezy (multi+reCV) & 13.58 (13.59)     \\
6. ecpcEN squeezy        & 5.48 (3.01)       & 3. fwEN (continuous)     & 14.11 (11.13)     \\
9. squeezy (single+reCV) & 7.45 (5.77)       & 5. gren                  & 351.52 (96.74)    \\
8. squeezy (multi)       & 10.47 (10.50)     & 4. ipf                   & 422.39 (259.80)  \\\hline
\end{tabular}%
}
\caption{Results of 10-fold CV in miRNA data example. Mean time per fold and standard deviation over folds, sorted from shortest to longest time.}
\label{tab:miRNATime}
\end{table}

\section{Discussion}\label{par:discussion}
%extension: MML.noDeriv (mgcv package optimisation)
%extension: wood also provides second derivatives, may be beneficial for stabilising the optimisation
%generalisability advantage tov gren: ook naar andere modellen. hier voor glms, maar partial likelihoods zouden ook moeten kunnen.
The proposed method, termed \texttt{squeezy}, computes fast marginal likelihood estimates for group-adaptive elastic net penalties. 
The method estimates those penalties by deducing them from group-adaptive ridge penalties, for
which we derive a low-dimensional representation of the Taylor approximation of the marginal
likelihood and its first derivative for generalised linear models, using results from \citep{wood2011fast}.
\texttt{Squeezy} implements this and uses Nelder-Mead optimisation to find the penalties. An extension would be to also use the hessian \citep{wood2011fast}, which may speed up the optimisation.

An alternative implementation of our method uses the R-package \texttt{mgcv} \citep{wood2011fast} to estimate the ridge penalties from the linear predictors, as detailed in \citep{van2020fast}. 
As \texttt{mgcv} does not allow for more variables than samples, this solution cannot include unpenalised variables. 
A fix to this is to include these as pre-estimated offsets. 
In the absence of unpenalised variables, we found that the two implementations provided similar results at comparable computing times. The alternative solution opens up for application of \texttt{squeezy} to a wider variety of models (provided that the elastic net counter part is also available) such as the penalised Cox model for survival data. 

%extension: other priors or hierarchical priors for shrinkage or selection on group level
We showed that the marginal likelihood of ridge models also approximates the marginal likelihood of elastic net models, by using the asymptotic multivariate normality of the linear predictor. This result also holds for other priors with finite variance. When co-data includes many groups, for example, a hierarchical prior could be used for shrinkage or selection on group level. The result does not hold for priors with infinite variance, such as the highly sparse horseshoe prior \citep{carvalho2009handling}. Moreover, the practical use of our method for other relatively sparse priors should be tested. 

The method transforms the ridge penalties to elastic net penalties using the variance function of the elastic net prior. We showed in a data example that it is beneficial to recalibrate the transformed elastic net penalties in logistic regression.
Furthermore, the simulation study and data example showed that our method is more scalable and faster than other group-adaptive methods, while it benefits from co-data and performs as well as or better than other (group-adaptive) methods. 

Should squeezy perform inferior to other methods, it may be useful to check whether this might be due to invalidity of the multivariate normal assumption for $\bs{\eta} = X\bs{\beta}$. Once the elastic net prior(s) for $\bs{\beta}$ are known, it is straightforward to generate multiple $\bs{\eta}$ realisations. Then, we suggest using those to generate a chi-square-based Q-Q plot, e.g. by the \texttt{R}-package \texttt{MVN} \citep{korkmaz2014mvn}, to verify multivariate normality.

%extension: AIC
As our method is based on (marginal) likelihood, it in principle facilitates model selection in terms of the hyper-parameters (e.g. single group versus multi-group) by the use of information criteria \citep{greven2010behaviour}. To what extent which criterion is useful in our setting is a topic for further research.

%software
We provide the \texttt{R}-package \texttt{squeezy} and scripts demonstrating the package and reproducing the results on
\url{https://github.com/Mirrelijn/squeezy}.

%%%%%%%%%%%%%%%%%%%%%%%%%%%%%%%%%%%%%%%%%%%%%%
%% Support information (funding), if any,   %%
%% should be provided in the                %%
%% Acknowledgements section.                %%
%%%%%%%%%%%%%%%%%%%%%%%%%%%%%%%%%%%%%%%%%%%%%%
\section*{Acknowledgements}
The first author is supported by ZonMw TOP grant COMPUTE CANCER (40-
00812-98-16012). The authors would like to thank Lodewyk Wessels and Soufiane Mourragui (Netherlands Cancer Insitute) for the many worthwhile discussions.


%\printbibliography
{\small
\bibliography{BIB} }

\begin{thebibliography}{20}
\providecommand{\natexlab}[1]{#1}
\providecommand{\url}[1]{\texttt{#1}}
\expandafter\ifx\csname urlstyle\endcsname\relax
  \providecommand{\doi}[1]{doi: #1}\else
  \providecommand{\doi}{doi: \begingroup \urlstyle{rm}\Url}\fi

\bibitem[Boulesteix et~al.(2017)Boulesteix, De~Bin, Jiang, and
  Fuchs]{boulesteix2017ipf}
Anne-Laure Boulesteix, Riccardo De~Bin, Xiaoyu Jiang, and Mathias Fuchs.
\newblock Ipf-lasso: Integrative-penalized regression with penalty factors for
  prediction based on multi-omics data.
\newblock \emph{Computational and mathematical methods in medicine}, 2017,
  2017.

\bibitem[Carvalho et~al.(2009)Carvalho, Polson, and
  Scott]{carvalho2009handling}
Carlos~M Carvalho, Nicholas~G Polson, and James~G Scott.
\newblock Handling sparsity via the horseshoe.
\newblock In \emph{Artificial Intelligence and Statistics}, pages 73--80.
  AISTATS, 2009.

\bibitem[Eicker(1966)]{eicker1966multivariate}
F~Eicker.
\newblock A multivariate central limit theorem for random linear vector forms.
\newblock \emph{The Annals of Mathematical Statistics}, pages 1825--1828, 1966.

\bibitem[Friedman et~al.(2010)Friedman, Hastie, and
  Tibshirani]{friedman2010regularization}
Jerome Friedman, Trevor Hastie, and Rob Tibshirani.
\newblock Regularization paths for generalized linear models via coordinate
  descent.
\newblock \emph{Journal of statistical software}, 33\penalty0 (1):\penalty0 1,
  2010.

\bibitem[Greven and Kneib(2010)]{greven2010behaviour}
Sonja Greven and Thomas Kneib.
\newblock On the behaviour of marginal and conditional aic in linear mixed
  models.
\newblock \emph{Biometrika}, 97\penalty0 (4):\penalty0 773--789, 2010.

\bibitem[Jacob et~al.(2009)Jacob, Obozinski, and Vert]{jacob2009group}
Laurent Jacob, Guillaume Obozinski, and Jean-Philippe Vert.
\newblock Group lasso with overlap and graph lasso.
\newblock In \emph{Proceedings of the 26th annual international conference on
  machine learning}, pages 433--440. ACM, 2009.

\bibitem[Korkmaz et~al.(2014)Korkmaz, Goksuluk, and Zararsiz]{korkmaz2014mvn}
Selcuk Korkmaz, Dincer Goksuluk, and Gokmen Zararsiz.
\newblock Mvn: An r package for assessing multivariate normality.
\newblock \emph{The R Journal}, 6\penalty0 (2):\penalty0 151--162, 2014.

\bibitem[Meier et~al.(2008)Meier, van~de Geer, and B{\"u}hlmann]{Meier2008}
L.~Meier, S.~van~de Geer, and P.~B{\"u}hlmann.
\newblock The group {L}asso for logistic regression.
\newblock \emph{J. R. Stat. Soc. Ser. B Stat. Methodol.}, 70\penalty0
  (1):\penalty0 53--71, 2008.
\newblock ISSN 1369-7412.

\bibitem[M{\"u}nch et~al.(2019)M{\"u}nch, Peeters, van~der Vaart, and van~de
  Wiel]{munch2018adaptive}
Magnus~M M{\"u}nch, Carel~FW Peeters, Aad~W van~der Vaart, and Mark~A van~de
  Wiel.
\newblock Adaptive group-regularized logistic elastic net regression.
\newblock \emph{Biostatistics}, 12 2019.
\newblock ISSN 1465-4644.
\newblock \doi{10.1093/biostatistics/kxz062}.
\newblock kxz062.

\bibitem[Neerincx et~al.(2015)Neerincx, Sie, Van De~Wiel, Van~Grieken,
  Burggraaf, Dekker, Eijk, Ylstra, Verhoef, Meijer, et~al.]{neerincx2015mir}
Maarten Neerincx, DLS Sie, MA~Van De~Wiel, NCT Van~Grieken, JD~Burggraaf,
  H~Dekker, PP~Eijk, Bauke Ylstra, C~Verhoef, GA~Meijer, et~al.
\newblock Mir expression profiles of paired primary colorectal cancer and
  metastases by next-generation sequencing.
\newblock \emph{Oncogenesis}, 4\penalty0 (10):\penalty0 e170, 2015.

\bibitem[Neerincx et~al.(2018)Neerincx, Poel, Sie, van Grieken, Shankaraiah,
  Van Der Wolf-De, van Waesberghe, Burggraaf, Eijk, Verhoef,
  et~al.]{neerincx2018combination}
Maarten Neerincx, Dennis Poel, Daoud~LS Sie, Nicole~CT van Grieken, Ram~C
  Shankaraiah, Floor~SW Van Der Wolf-De, Jan-Hein~TM van Waesberghe, Jan-Dirk
  Burggraaf, Paul~P Eijk, Cornelis Verhoef, et~al.
\newblock Combination of a six microrna expression profile with four
  clinicopathological factors for response prediction of systemic treatment in
  patients with advanced colorectal cancer.
\newblock \emph{PloS one}, 13\penalty0 (8):\penalty0 e0201809, 2018.

\bibitem[Tay et~al.(2020)Tay, Aghaeepour, Hastie, and
  Tibshirani]{tay2020feature}
J~Kenneth Tay, Nima Aghaeepour, Trevor Hastie, and Robert Tibshirani.
\newblock Feature-weighted elastic net: using" features of features" for better
  prediction.
\newblock \emph{arXiv preprint arXiv:2006.01395}, 2020.

\bibitem[Tomczak et~al.(2015)Tomczak, Czerwi{\'n}ska, and
  Wiznerowicz]{tomczak2015cancer}
Katarzyna Tomczak, Patrycja Czerwi{\'n}ska, and Maciej Wiznerowicz.
\newblock The cancer genome atlas (tcga): an immeasurable source of knowledge.
\newblock \emph{Contemporary oncology}, 19\penalty0 (1A):\penalty0 A68, 2015.

\bibitem[van~de Wiel et~al.(2019)van~de Wiel, Te~Beest, and
  M{\"u}nch]{van2019learning}
Mark~A van~de Wiel, Dennis~E Te~Beest, and Magnus~M M{\"u}nch.
\newblock Learning from a lot: Empirical bayes for high-dimensional model-based
  prediction.
\newblock \emph{Scandinavian Journal of Statistics}, 46\penalty0 (1):\penalty0
  2--25, 2019.

\bibitem[van~de Wiel et~al.(2020)van~de Wiel, van Nee, and
  Rauschenberger]{van2020fast}
Mark~A van~de Wiel, Mirrelijn~M van Nee, and Armin Rauschenberger.
\newblock Fast cross-validation for multi-penalty ridge regression.
\newblock \emph{arXiv preprint arXiv:2005.09301}, 2020.

\bibitem[{van Nee} et~al.(2020){van Nee}, Wessels, and {van de
  Wiel}]{vanNee2020flexible}
Mirrelijn~M {van Nee}, Lodewyk~FA Wessels, and Mark~A {van de Wiel}.
\newblock Flexible co-data learning for high-dimensional prediction.
\newblock \emph{arXiv preprint arXiv:2005.04010}, 2020.

\bibitem[Veerman et~al.(2019)Veerman, Leday, and van~de
  Wiel]{veerman2019estimation}
Jurre~R Veerman, Gwena{\"e}l~GR Leday, and Mark~A van~de Wiel.
\newblock Estimation of variance components, heritability and the ridge penalty
  in high-dimensional generalized linear models.
\newblock \emph{Communications in Statistics-Simulation and Computation}, pages
  1--19, 2019.

\bibitem[Wood(2008)]{wood2008fast}
Simon~N Wood.
\newblock Fast stable direct fitting and smoothness selection for generalized
  additive models.
\newblock \emph{Journal of the Royal Statistical Society: Series B (Statistical
  Methodology)}, 70\penalty0 (3):\penalty0 495--518, 2008.

\bibitem[Wood(2011)]{wood2011fast}
Simon~N Wood.
\newblock Fast stable restricted maximum likelihood and marginal likelihood
  estimation of semiparametric generalized linear models.
\newblock \emph{Journal of the Royal Statistical Society: Series B (Statistical
  Methodology)}, 73\penalty0 (1):\penalty0 3--36, 2011.

\bibitem[Zou and Hastie(2005)]{zou2005regularization}
Hui Zou and Trevor Hastie.
\newblock Regularization and variable selection via the elastic net.
\newblock \emph{Journal of the Royal Statistical Society: Series B (Statistical
  Methodology)}, 67\penalty0 (2):\penalty0 301--320, 2005.

\end{thebibliography}

%%%%%%%%%%%%%%%%%%%%%%%%%%%%%%%%%%%%%%%%%%%%%%
%% Single Appendix:                         %%
%%%%%%%%%%%%%%%%%%%%%%%%%%%%%%%%%%%%%%%%%%%%%%
%\begin{appendix}
%\section*{???}%% if no title is needed, leave empty \section*{}.
%\end{appendix}
%%%%%%%%%%%%%%%%%%%%%%%%%%%%%%%%%%%%%%%%%%%%%%
%% Multiple Appendixes:                     %%
%%%%%%%%%%%%%%%%%%%%%%%%%%%%%%%%%%%%%%%%%%%%%%
\newpage

\begin{appendix}
\section*{Appendix}
\renewcommand{\appendixname}{}
\setcounter{section}{0}
\setcounter{table}{0}
\renewcommand{\thetable}{\thesection\arabic{table}}%
\setcounter{figure}{0}
\renewcommand{\thefigure}{\thesection\arabic{figure}}%
\setcounter{equation}{0}
\renewcommand{\theequation}{\thesection\arabic{equation}}

\section{Details for Laplace approximate marginal likelihood for group-adaptive ridge models}\label{ap:method}
%Consider generalised linear models with group-adaptive ridge prior as described in Section \ref{par:method}. 
We provide details of the derivation of a low-dimensional representation of the Laplace approximation for the marginal likelihood and its first derivative as stated in Equations \eqref{eq:MML} and \eqref{eq:partiallam}. 
When some covariates are left unpenalised, we need to integrate over the penalised covariates only. \cite{wood2011fast} forms an ortogonal basis $U$ for the range space of $S$. For group-adaptive ridge penalties, this orthogonal basis is simply given by the identity matrix, as $S$ is diagonal with zeros for unpenalised covariates. The proposed reparametrisation of $X$ and $S$ boils down to simply taking the columns in $X$ that correspond to the penalised variables, $\bar{X}=X_{pen}$, and taking the columns and rows of the penalty matrix that correspond to the penalised variables, $\bar{S}=S_{pen}=\Lambda_{pen}=\sum_g\lambda_gS_{g,pen}$.

\subsection{Laplace approximate marginal likelihood}
The high-dimensional log-marginal likelihood approximation is given in Equation (5) in \citep{wood2011fast} and may be written as:
\begin{align}]\label{eq:MMLhigh}
    -\ell(\bs{\lambda}) &\approx \frac{D_p}{2\phi} - \ell_s(\phi) + \frac{\log(\bar{X}^TW\bar{X}+\bar{S}) - \log(|S|_+)}{2}\nonumber\\
    &= -\ell(\hat{\bs{\beta}}) + \frac{1}{2\phi}\hat{\bs{\beta}}^T\Lambda_{pen}\hat{\bs{\beta}} +\frac{1}{2}\log(|X_{pen}^TWX_{pen}+\Lambda_{pen}|) -\frac{1}{2}\log(|\Lambda_{pen}|).
    % &= -\ell(\hat{\bs{\beta}}) + \frac{1}{2\phi}\hat{\bs{\beta}}^T\phi\Lambda_{pen}\hat{\bs{\beta}} -\frac{1}{2}\log(|(X_{pen}^TWX_{pen}+\phi\Lambda_{pen})^{-1}\phi\Lambda_{pen}|)
\end{align}
When we use Newton iterations to obtain $\hat{\bs{\beta}}$ or $\hat{\bs{\eta}}$, upon convergence:
\begin{align*}
    \hat{\bs{\beta}}=(X^TWX+\Lambda)^{-1}X^T(\bs{y}-\bs{\mu}+WX\hat{\bs{\beta}}),\\
    \hat{\bs{\eta}}=X(X^TWX+\Lambda)^{-1}X^T(\bs{y}-\bs{\mu}+W\hat{\bs{\eta}}).
\end{align*}
We substitute this in the second term of the log-marginal likelihood approximation:
\begin{align*}
    \hat{\bs{\beta}}^T\Lambda\hat{\bs{\beta}}&= (\bs{y}-\bs{\mu}+WX\hat{\bs{\beta}})^TX(X^TWX+\Lambda)^{-1}\Lambda\hat{\bs{\beta}}\\
    &= (\bs{y}-\bs{\mu}+WX\hat{\bs{\beta}})^TX(I_{p\times p} - (X^TWX+\Lambda)^{-1}X^TWX)\hat{\bs{\beta}}\\
    &= (\bs{y}-\bs{\mu}+WX\hat{\bs{\beta}})^T\hat{\bs{\eta}} - \hat{\bs{\eta}}^TW\hat{\bs{\eta}}\\
    &= (\bs{y}-\bs{\mu})^T\hat{\bs{\eta}}.
\end{align*}
The latter two terms may be rewritten in a lower-dimensional representation by:
\begin{align*}
    &\frac{1}{2}\log(|X_{pen}^TWX_{pen}+\Lambda_{pen}|) -\frac{1}{2}\log(|\Lambda_{pen}|)
    =-\frac{1}{2}\log(|(X_{pen}^TWX_{pen}+\Lambda_{pen})^{-1}\Lambda_{pen}|)\\
    \qquad &= -\frac{1}{2}\log(|I_{pen\times pen} - (X_{pen}^TWX_{pen}+\Lambda_{pen})^{-1}X_{pen}^TWX_{pen}|)\\
    \qquad &= -\frac{1}{2}\log(|I_{n\times n} - WX_{pen}(X_{pen}^TWX_{pen}+\Lambda_{pen})^{-1}X_{pen}^T|)\\
    \qquad &= -\frac{1}{2}\log(|I_{n\times n} - WH_{pen}|),
\end{align*}
where efficient representations for $H_{pen}$ exist \citep{van2020fast}.

So, the low-dimensional Laplace approximation for the minus log-marginal likelihood is:
\begin{align*}
    -\ell(\bs{\lambda})&\approx -\ell(\hat{\bs{\eta}}) + \frac{1}{2\phi}(\bs{y}-\bs{\mu})^T\hat{\bs{\eta}} -\frac{1}{2}\log(|I_{n\times n} - WH_{pen}|).
\end{align*}

\subsection{First derivative of Laplace approximate marginal likelihood}
We derive low-dimensional representations of the derivatives to $\rho_g=\log(\lambda_g)$ for $g=1,..,G$, as derived in a high-dimensional representation by \citep{wood2011fast}.

\subsubsection{Derivative of \texorpdfstring{$\frac{\partial \hat{\bs{\eta}}}{\partial \rho_g}$}{detadrho}}
Denote by $I_g\in\mathbb{R}^{p\times p}$ the diagonal matrix with diagonal element $k$ equal to $1$ if $k$ is in group $g$ and $0$ otherwise and define $\Lambda_g=\lambda_gI_g$. The derivative for $\hat{\bs{\beta}}$ is then given by \citep{wood2011fast}:
\begin{align*}
    \frac{\partial \hat{\bs{\beta}}}{\partial \rho_g}
    &= -\exp(\rho_g)(X^TWX+S)^{-1}S_g\hat{\bs{\beta}}\\
    &= -\exp(\rho_g)(X^TWX+\Lambda)^{-1}\Lambda_g\frac{1}{\lambda_g}\hat{\bs{\beta}}\\
    &= -\exp(\rho_g)(X^TWX+\Lambda)^{-1}\Lambda \frac{1}{\lambda_g}I_g\hat{\bs{\beta}}\\
    &= -\exp(\rho_g)\left[I_{p\times p} - (X^TWX+\Lambda)^{-1}X^TWX\right] \frac{1}{\lambda_g}I_g\hat{\bs{\beta}}\\
    &= -\exp(\rho_g)\left[I_{p\times p} - (X^TWX+\Lambda)^{-1}X^TWX\right] \frac{1}{\lambda_g}I_g (X^TWX+\Lambda)^{-1}X^T\left(\bs{y}-\bs{\mu}+W\hat{\bs{\eta}}\right).
\end{align*}
We can write the derivative for $\hat{\bs{\eta}}$ as:
\begin{align*}
    \frac{\partial \hat{\bs{\eta}}}{\partial \rho_g}
    &= X\frac{\partial \hat{\bs{\beta}}}{\partial \rho_g}\\
    &= -\exp(\rho_g)X\left[I_{p\times p} - (X^TWX+\Lambda)^{-1}X^TWX\right] \frac{1}{\lambda_g}I_g (X^TWX+\Lambda)^{-1}X^T\left(\bs{y}-\bs{\mu}+W\hat{\bs{\eta}}\right)\\
    &= -\exp(\rho_g)\left[I_{n\times n} - X(X^TWX+\Lambda)^{-1}X^TW\right] X\frac{1}{\lambda_g}I_g (X^TWX+\Lambda)^{-1}X^T\left(\bs{y}-\bs{\mu}+W\hat{\bs{\eta}}\right)\\
    &= -\exp(\rho_g)\left[I_{n\times n} - HW\right] \frac{1}{\lambda_g}H_g^T\left(\bs{y}-\bs{\mu}+W\hat{\bs{\eta}}\right),
\end{align*}
with $H_g$, which can be seen as a contribution of the $g^{th}$ group to the hat matrix, defined as:
\begin{align*}
\begin{split}
    &H_g:=X(X^TWX+\Lambda)^{-1}I_gX^T
    = W^{-\frac{1}{2}}P_1W^{\frac{1}{2}}\left(I_{n\times n} - X_{pen}\Lambda^{-1}_{pen}X_{pen}^T W^{\frac{1}{2}}P_1W^{-\frac{1}{2}} \right.\\
    &\qquad \cdot\left.[W^{-1}+X_{pen}\Lambda^{-1}X_{pen}^TW^{\frac{1}{2}}P_1W^{-\frac{1}{2}}]^{-1}\right)\lambda_g^{-1}X_gX_g^T,
\end{split}
\end{align*}
where we have used \citep{van2020fast}:
\begin{align*}
&P_1=I_{n\times n} - W^{\frac{1}{2}}X_{unpen}(X_{unpen}^TWX_{unpen})^{-1}X_{unpen}^TW^{\frac{1}{2}},\\
\begin{split}
    &[X(X^TWX+\Lambda)^{-1}]_{pen} = W^{-\frac{1}{2}}P_1W^{\frac{1}{2}}\left(X_{pen}\Lambda^{-1}_{pen} - X_{pen}\Lambda^{-1}_{pen}X_{pen}^T W^{\frac{1}{2}}P_1W^{-\frac{1}{2}} \right.\\
    &\qquad \cdot\left.[W^{-1}+X_{pen}\Lambda^{-1}X_{pen}^TW^{\frac{1}{2}}P_1W^{-\frac{1}{2}}]^{-1}X_{pen}\Lambda_{pen}^{-1} \right).
\end{split}
\end{align*}

\subsubsection{Derivative of \texorpdfstring{$\frac{\partial \ell(\hat{\bs{\eta}})}{\partial \rho_g}$}{delldrho}}
\cite{wood2008fast} derives partials for the deviance of generalised linear models:
\begin{align*}
    \frac{\partial \ell(\hat{\bs{\beta}})}{\partial \hat{\bs{\beta}}} &= \frac{1}{\phi} X^T\frac{\bs{y}-\bs{\mu}}{V(\bs{\mu})\odot g'(\bs{\mu})},
\end{align*}
where division by $V(\bs{\mu})\odot  g'(\bs{\mu})$ is element-wise and $\odot$ representing element-wise multiplication.
The derivative of the log likelihood is then given by:
\begin{align*}
    \frac{\partial \ell(\hat{\bs{\eta}})}{\partial \rho_g}=\frac{\partial \ell(\hat{\bs{\beta}})}{\partial \rho_g} &= \frac{\partial \ell(\hat{\bs{\beta}})}{\partial \hat{\bs{\beta}}}^T \frac{\partial \hat{\bs{\beta}}}{\partial \rho_g}\nonumber\\
    &= -\exp(\rho_g)\frac{1}{\phi} \left(\frac{\bs{y}-\bs{\mu}}{V(\bs{\mu})}\right)^TG'^{-1}X (X^TWX+\Lambda)^{-1}\Lambda_g\frac{1}{\lambda_g}\hat{\bs{\beta}}\nonumber\\
    &= -\exp(\rho_g)\frac{1}{\phi} \left(\frac{\bs{y}-\bs{\mu}}{V(\bs{\mu})}\right)^TG'^{-1} \left[I_{n\times n} - HW\right] \frac{1}{\lambda_g}H_g^T\left(\bs{y}-\bs{\mu}+W\hat{\bs{\eta}}\right),
\end{align*}
with $G'$ a diagonal matrix with diagonal elements $g'(\mu_i)$.

\subsubsection{Derivative of \texorpdfstring{$\frac{1}{2\phi}(\bs{y}-\bs{\mu})^T\hat{\bs{\eta}}$}{d2drho}}
We have for the low-dimensional penalty term, recall that $\bs{\mu}=g^{-1}(\bs{\eta})$:
\begin{align*}
    \frac{\partial }{\partial \rho_g} \left(\frac{1}{2\phi}(\bs{y}-\bs{\mu})^T\hat{\bs{\eta}}\right) &= \frac{1}{2\phi} \left(\frac{\partial (\bs{y}-\bs{\mu})^T}{\partial \rho_g}\hat{\bs{\eta}} + (\bs{y}-\bs{\mu})^T \frac{\partial \hat{\bs{\eta}}}{\partial \rho_g}\right) \nonumber\\
    &= \frac{1}{2\phi} \left(\frac{\partial (-\bs{\mu}^T)}{\partial \rho_g}\hat{\bs{\eta}} + (\bs{y}-\bs{\mu})^T \frac{\partial \hat{\bs{\eta}}}{\partial \rho_g}\right) \nonumber\\
    &=\frac{1}{2\phi}\left( -\frac{\partial \hat{\bs{\eta}}^T}{\partial \rho_g}G'^{-1}\hat{\bs{\eta}} + (\bs{y}-\bs{\mu})^T \frac{\partial \hat{\bs{\eta}}}{\partial \rho_g}\right) \nonumber\\
    &=\frac{1}{2\phi}\left(-G'^{-1}\hat{\bs{\eta}} + \bs{y}-\bs{\mu} \right)^T\frac{\partial \hat{\bs{\eta}}}{\partial \rho_g} \nonumber\\
    &=-\exp(\rho_g)\frac{1}{2\phi}\left(-G'^{-1}\hat{\bs{\eta}} + \bs{y}-\bs{\mu} \right)^T\left[I_{n\times n} - HW\right] \frac{1}{\lambda_g}H_g^T\left(\bs{y}-\bs{\mu}+W\hat{\bs{\eta}}\right).
\end{align*}

\subsubsection{Derivative of \texorpdfstring{$-\frac{1}{2}\log(|\bs{I}_{n\times n} - \bs{W}\bs{H}_{pen}|)$}{dlastdrho}}
We have for the third term in Equation \eqref{eq:MMLhigh}:
\begin{align*}
    \frac{\partial}{\partial \rho_g}\left(\frac{1}{2}\log(|X_{pen}^TWX_{pen}+\Lambda_{pen}|\right)
    &= \frac{1}{2}tr\left((X_{pen}^TWX_{pen}+\Lambda_{pen})^{-1}\frac{\partial X_{pen}^TWX_{pen}+\Lambda_{pen}}{\partial\rho_g} \right)\\
    &= \frac{1}{2}tr\left((X_{pen}^TWX_{pen}+\Lambda_{pen})^{-1}X_{pen}^T\frac{\partial W}{\partial\rho_g}X_{pen} \right)\\
    &\qquad + \frac{1}{2}tr\left((X_{pen}^TWX_{pen}+\Lambda_{pen})^{-1} \Lambda_g \right)\\
    &= \frac{1}{2}tr\left(X_{pen}(X_{pen}^TWX_{pen}+\Lambda_{pen})^{-1}X_{pen}^T\frac{\partial W}{\partial\rho_g} \right)\\
    &\qquad + \frac{1}{2}tr\left((X_{pen}^TWX_{pen}+\Lambda_{pen})^{-1} \Lambda_g \right)\\
    &= \frac{1}{2}tr\left( H_{pen}\frac{\partial W}{\partial\rho_g}\right) + \frac{1}{2}tr(I_g)\\
    &\qquad - \frac{1}{2}tr\left(\Lambda_{pen}^{-1}X_{pen}^T(W^{-1}+X_{pen}\Lambda_{pen}^{-1}X_{pen}^T)^{-1}X_{pen}I_g\right)\\
    &= \frac{1}{2}tr\left( H_{pen}\frac{\partial W}{\partial\rho_g}\right) + \frac{1}{2}tr(I_g)\\
    &\qquad - \frac{1}{2}tr\left(\lambda_g^{-1}X_gX_g^T(W^{-1}+X_{pen}\Lambda_{pen}^{-1}X_{pen}^T)^{-1}\right),
\end{align*}
where we have used the following equalities:
\begin{align*}
    (X_{pen}^TWX_{pen}+\Lambda_{pen})^{-1}&=\Lambda_{pen}^{-1} - \Lambda_{pen}^{-1}X_{pen}^T(W^{-1}+X_{pen}\Lambda_{pen}^{-1}X_{pen}^T)^{-1}X_{pen}\Lambda_{pen}^{-1}\\
    H_{pen} &= X_{pen}(X_{pen}^TWX_{pen}+\Lambda_{pen})^{-1}X_{pen}^T,
\end{align*}
with the partials of $W\in\mathbb{R}^{n\times n}$ readily obtained from \citep{wood2011fast}. 

For the latter term in Equation \eqref{eq:MMLhigh}:
\begin{align*}
    \frac{\partial}{\partial \rho_g}\left(-\frac{1}{2}\log(|\Lambda_{pen}|)\right)&=-\frac{1}{2} tr\left(\Lambda_{pen}^{-1} \frac{\partial\Lambda_{pen}}{\partial \rho_g}\right)\nonumber\\
    &=-\frac{1}{2} tr\left(\Lambda_{pen}^{-1}I_g\frac{\partial\lambda_g}{\partial\rho_g}\right)\nonumber\\
    &=-\frac{1}{2}tr(I_g)\frac{1}{\lambda_g}\lambda_g\nonumber\\
    &=-\frac{1}{2}tr(I_g).
\end{align*}

So, the derivative of the Laplace approximate minus log marginal likelihood is given by:
\begin{align*}
   \frac{\partial -\ell(\bs{\rho})}{\partial \rho_g} &= \frac{1}{\phi} \left(\frac{\bs{y}-\bs{\mu}}{V(\bs{\mu})}\right)^TG'^{-1} \left[I_{n\times n} - HW\right] H_g^T\left(\bs{y}-\bs{\mu}+W\hat{\bs{\eta}}\right)\\
   &\qquad -\frac{1}{2\phi}\left(-G'^{-1}\hat{\bs{\eta}} + \bs{y}-\bs{\mu} \right)^T\left[I_{n\times n} - HW\right] H_g^T\left(\bs{y}-\bs{\mu}+W\hat{\bs{\eta}}\right)\\
   &\qquad +\frac{1}{2}tr\left( H_{pen}\frac{\partial W}{\partial\rho_g}\right)  - \frac{1}{2}tr\left(\lambda_g^{-1}X_gX_g^T(W^{-1}+X_{pen}\Lambda_{pen}^{-1}X_{pen}^T)^{-1}\right).
\end{align*}

\subsection{Details for linear regression}
Denote by $\phi(\cdot;\mu,\sigma^2)$ the normal pdf with mean $\mu$ and variance $\sigma^2$.
The Laplace approximation is in fact exact for linear regression. We need the following parameters:
\begin{align*}
    \phi&=\sigma^2,
    V(\mu)=1,
    \ell(\hat{\bs{\eta}})=\sum_{i=1}^n\log(\phi(y_i;\hat{\bs{\eta}}_i,\sigma^2)),\\
    \bs{\mu}&=\hat{\bs{\eta}},\ W=I_{n\times n},\ g(\mu)=\mu,\ g'(\mu)=1,\ G'^{-1}=I_{n\times n}.
\end{align*}
The partial derivative of the minus log likelihood to the scale parameter $\phi=\sigma^2$ from Equation \eqref{eq:partialphi} is given by:
\begin{align*}
    \frac{\partial -\ell(\hat{\bs{\eta}},\sigma^2)}{\partial \sigma^2} &= \frac{n}{2\sigma^2} - \frac{1}{2\sigma^4}(\bs{y}-\hat{\bs{\eta}})^T(\bs{y}-\hat{\bs{\eta}}).
\end{align*}

\subsection{Details for logistic regression}
We need the following parameters:
\begin{align*}
    \phi&=1,
    V(\mu_i)=\mu_i(1-\mu_i),
    \ell(\hat{\bs{\eta}})=\sum_{i=1}^ny_i\log(expit(\hat{\eta}_i)) + (1-y_i)\log(1-expit(\hat{\eta}_i)),\\
    \bs{\mu}&=expit(\hat{\bs{\eta}})=(1+\exp(-\hat{\bs{\eta}}))^{-1},\\
    W&=diag(\bs{\mu}\odot(1-\bs{\mu})),\ \alpha_i=1,\ \frac{\partial w_i}{\partial \eta_i}=\mu_i(1-\mu_i)(2\mu_i-1),\\
    g(\mu_i)&=logit(\mu_i),\ g'(\mu_i)=(\mu_i(1-\mu_i))^{-1},\ G'^{-1}=W,\ g''(\mu_i)=\frac{2\mu_i-1}{\mu_i^2(1-\mu_i)^2}.
\end{align*}

\section{Additional figures to the data examples}\label{ap:dataexamples}
\begin{figure}
    \centering
    \includegraphics[width=\textwidth]{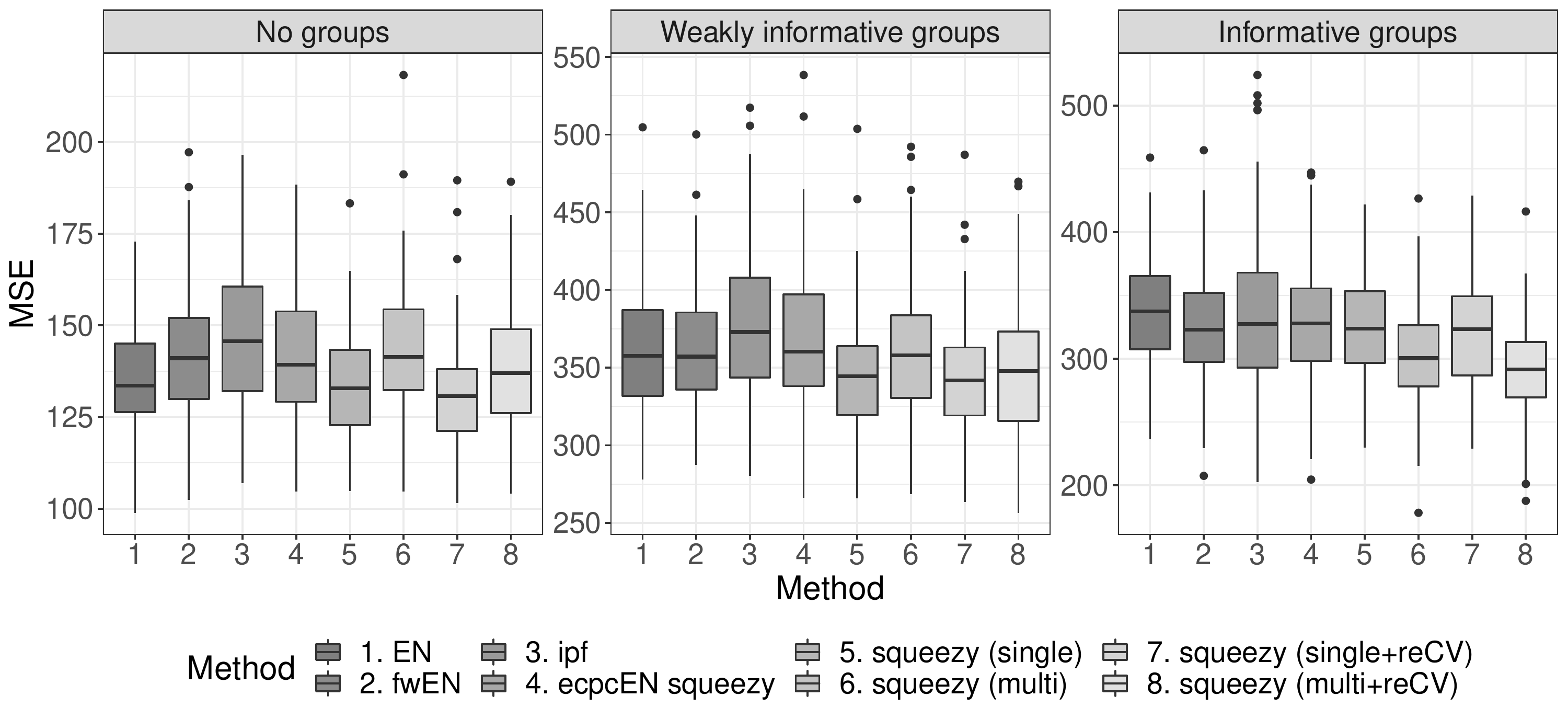}
    \caption{Model-based simulation study. Box plots of the MSE performance on the test sets of $50$ pairs of training and test sets, $\alpha=0.8$.}
    \label{fig:SimMSE08}
\end{figure}

Figure \ref{fig:SimMSE08} shows the MSE performance in the model-based simulation study for $\alpha=0.8$.

\end{appendix}

\end{document}